# Generating an ATL Model Checker using an Attribute Grammar

FLORIN STOICA[1], LAURA FLORENTINA STOICA
Department of Computer Science
"Lucian Blaga" University of Sibiu, Faculty of Sciences
Str. Dr. Ion Ratiu 5-7, 550012, Sibiu
ROMANIA
florin.stoica@ulbsibiu.ro, laura.cacovean@ulbsibiu.ro

*Abstract:* In this paper we use attribute grammars as a formal approach for model checkers development. Our aim is to design an ATL (Alternating-Time Temporal Logic) model checker from a context-free grammar which generates the language of the ATL formulas. An attribute grammar may be informally defined as a context-free grammar which is extended with a set of attributes and a collection of semantic rules. We use an ATL attribute grammar for specifying an operational semantics of the language of the ATL formulas by defining a translation into the language which describes the set of states from the ATL model where the corresponding ATL formulas are satisfied.
We provide a formal definition for an attribute grammar used as input for Another Tool for Language Recognition (ANTLR) to generate an ATL model checker. Also, the technique of implementing the semantic actions in ANTLR is presented, which is the concept of connection between attribute evaluation in the grammar that generates the language of ATL formulas and algebraic compiler implementation that represents the ATL model checker. The original implementation of the model checking algorithm is based on Relational Databases and Web Services. Several database systems and Web Services technologies were used for evaluating the system performance in verification of large ATL models.

## 1 Introduction

Model checking is a technology widely used for the automated system verification and represents a technique for verifying that finite state systems satisfy specifications expressed in the language of temporal logics.

In this paper we present a new ATL model checking tool used for verification of open systems. An open system interacts with its environment and its behavior depends on the state of the system as well as the behavior of the environment.

The model checking problem for ATL is to determine whether a given model satisfies a given specification expressed by an ATL formula.

Two most common methods of performing model checking are explicit enumeration of states of the model and respectively the use of symbolic methods.

Symbolic model checkers analyse the state space symbolically using Ordered Binary Decision Diagrams (OBDDs), which were introduced in [4]. A symbolic model checker represents the structure of the model itself symbolically using OBDDs to represent transition relations by Boolean expressions. The key to symbolic model checking is to perform all calculations directly using these Boolean expressions, rather than using the model structure explicitly. An efficient representation of the models using OBDDs can potentially allow much larger structures to be checked.

ATL has been implemented in several symbolic tools for the analysis of open systems such as MOCHA for the modular verification of heterogeneous systems [2] and MCMAS, a symbolic model checker specifically tailored to agent-based specifications and scenarios [11].

The aim of our research was to develop a reliable, easy to maintain, scalable model checker tool to improve applicability of ATL model checking in design of general-purpose computer software.

---
[1] Corresponding author

In the following we will present a short justification for the choice of the explicit-state model technique.

In [5] a comparison is presented between RULEBASE, a symbolic model checker developed at IBM Haifa Research Laboratory and the explicit LTL (Linear Temporal Logic) model checker SPIN [7]. The verified software was a distributed storage subsystem software application. The state space size handled by SPIN was $10^8$ in a 3-process model. Using symbolic model checking, RULEBASE keeps a compressed representation of the state space and thus was able to manage $10^{150}$ states. On the other hand, because of the limit on state size, RULEBASE could not represent a state large enough to include the information needed for more than 2-process configuration [5].

Most hardware designs are based on a clocked-approach and thus are synchronous. For these systems, the symbolic model checking approach is more appropriate [10].

On the other hand, for nondeterministic, high-level models of hardware protocols, it has previously been argued that explicit model checking is better than symbolic model checking [8]; this is because the communication mechanisms inherent in protocols tend to cause the BDDs in symbolic model checking to blow up [3].

In their basic form, symbolic approaches tend to perform poorly on asynchronous models where concurrent interleaving are the main source of explosion of the BDD representation, and explicit-state model-checkers have been the preferred approach for such models [3].

Concurrent software is asynchronous as the different components might be running on different processors or be interleaved by the scheduler of the operating system. Taking into account the above considerations, in our tool we are using an explicit-state model technique.

The most pressing challenge in model checking today is scalability [14]. A model-checking tool must be efficient, in terms of the size of the models it can reason about and the time and space it requires, in order to scaling its verification ability to handle real-world applications.

An orthogonal approach to increase the capacity of an explicit-state model checker tool is to exploit the memory and computational resources of multiple computers in a distributed computing environment [3]. Following this idea, our tool is based on Web Services technology to address the time constraints in verification of large models.

The process of verification of a ATL model requires defining a specification which is represented by an ATL formula, and then determining whether or not that specification it is satisfied in the model. Such a specification is performed using the ATL formulas language, which is based on well-established syntactic and semantic rules.

Verification of an ATL formula involves a translation of it, from the language in which it was defined to the language over the set of states of the model. The result of this translation will be the set of states that satisfy the given formula in the checked ATL model.

Most often the designing a translator is difficult to achieve and require significant efforts for construction and maintenance. There are now specialized tools that generate the full code required using a grammar specification of the source language.

Traditionally, the tools used for the two phases of the translation of the text, *lexical analysis* and *syntactic analysis* were LEX *(A Lexical Analyzer Generator)* and YACC *(Yet Another Compiler Compiler)*, or their GNU equivalent, FLEX, BISON, BYACC/J. The disadvantage of the tools LEX and YACC respectively FLEX and BISON is that they only generate *C* code and that code is not always easily understood by the user (complexity induced by the nature of analyzers they generate). BYACC / J is able to generate Java code, but supported semantic actions are rudimentary.

A high-performance analyzer generator is ANTLR [12] *(Another Tool for Language Recognition)*, capable of generating C++, C #, Java or Python code and represents the instrument used in this article. We will use Java as the target language into which will be developed our own ATL model checker tool.

The original contribution of our approach consists in development of the ATL model checker tool by designing an ANTLR attribute grammar upon which is then generated, using ANTLR, the entire model checker tool.

The evaluation of an ATL formula is done by automatic activation of the semantic actions associated with production rules in the process of parsing the respective ATL formula, supplied as input for the ATL model checker tool. The semantic actions are written in target language of the generated parser, in our case Java. These actions are incorporated in source code of the parser and are activated whenever the parser recognizes a valid syntactic construction in the translated ATL formula.

# 2 Material and methods

From a formal point of view, implementation of an ATL model checker will be accomplished through the implementation of an algebraic compiler $\mathcal{C}$ in two steps. First, we need a syntactic parser to verify the syntactic correctness of a formula φ. Second, we should deal with the semantics of the ATL language, respectively with the implementation of the ATL operators. By implementing these tasks, the algebraic compiler $\mathcal{C}$ translates an ATL formula φ to the set of states over which formula φ is satisfied.

In this paper we will present a model-checking algorithm based on procedure from [1]. Our original approach is represented by the generation of an ATL model checker from our specification grammar of ATL using ANTLR (Another Tool for Language Recognition) as parser generator. Also, we are using a database approach to implement the ATL temporal operators.

## 2.1 The ATL model

The Alternating-Time Temporal Logic (ATL) logic is interpreted over concurrent game structures, considered as natural models for compositions of open systems.

A *concurrent game structure* is defined as a tuple $S=\langle \Lambda, Q, \Gamma, \gamma, M, d, \delta \rangle$ with the following components: a nonempty finite set of all players $\Lambda = \{1, ..., k\}$; a finite set of *states* $Q$; a finite set of *propositions* (or *observables*) $\Gamma$; the **labelling** (or *observation*) function γ; a nonempty finite set of moves $M$; the **alternative moves** function $d$ and the **transition function** δ. For each state $q \in Q$, $\gamma(q) \subseteq \Gamma$ is the set of propositions *true* at $q$. For each player $a \in \{1,...,k\}$ and each state $q \in Q$, the alternative moves function $d: \Lambda \times Q \to \mathcal{P}(M)$ associates the set of available moves of player $a$ at state $q$, where $\mathcal{P}(M)$ denote the power-set of $M$. In the following, the set $d(a,q)$ will be denoted by $d_a(q)$. For each state $q \in Q$, a tuple $\langle j_1,...,j_k \rangle$ such that $j_a \in d_a(q)$ for each player $a \in \Lambda$, represents a *move vector* at $q$. The set of all potential move vectors at $q$ is denoted by $\overline{M}_q = d_1(q) \times ... \times d_k(q)$ and we denote by $\overline{M} = \bigcup_{q \in Q} \overline{M}_q$ the set of all potential move vectors within the concurrent game structure.

We define the **move function** $D: Q \to \mathcal{P}(\overline{M})$, such that $D(q) \subseteq \overline{M}_q$ is the set of move vectors at $q$. We write:

$$D_a = \bigcup_{q \in Q} d_a(q) \qquad (1)$$

for the set of available moves of player $a$ within the game structure $S$.

The transition function $\delta(q,j_1,...,j_k)$, associates to each state $q \in Q$ and each move vector $\langle j_1,...,j_k \rangle \in D(q)$ the state that results from state $q$ if every player $a \in \{1,...,k\}$ chooses move $j_a$.

A *computation* of $S$ is an infinite sequence $\lambda = q_0, q_1, ...$ such that $q_{i+1}$ is the successor of $q_i$, i.e. there is a move vector $\langle j_1,...,j_k \rangle \in D(q_i)$ such that $\delta(q_i, j_1,...,j_k) = q_{i+1}, \forall i \geq 0$ [1]. A *q-computation* is a computation starting at state $q$.

For a computation $\lambda$ and a position $i \geq 0$, we denote by $\lambda[i]$, $\lambda[0,i]$, and $\lambda[i,\infty]$ the $i$-th state of $\lambda$, the finite prefix $q_0, q_1, ..., q_i$ of $\lambda$, and the infinite suffix $q_i, q_{i+1} ...$ of $\lambda$, respectively [1].

## 2.2 ATL syntax and semantics

We denote by $\mathcal{F}_S(\mathcal{A})$ the set of all syntactically correct ATL formulas, defined over a concurrent game structure $S$ and a set of players $\mathcal{A} \subseteq \Lambda$.

Each formula from $\mathcal{F}_S(\mathcal{A})$ can be constructed using the following rules:

(R1) $\{p \mid p \in \Gamma\} \cup \{\tau, \bot\} \subseteq \mathcal{F}_S(\mathcal{A})$;

(R2) if $\{\varphi, \varphi_1, \varphi_2\} \subseteq \mathcal{F}_S(\mathcal{A})$ then $\{\neg \varphi, \varphi_1 \vee \varphi_2\} \subseteq \mathcal{F}_S(\mathcal{A})$;

*(R3)* if $\{\varphi, \varphi_1, \varphi_2\} \subseteq \mathcal{F}_S(\mathcal{A})$ then $\{\langle\langle\mathcal{A}\rangle\rangle \circ \varphi, \langle\langle\mathcal{A}\rangle\rangle \square \varphi, \langle\langle\mathcal{A}\rangle\rangle \varphi_1 U \varphi_2\} \subseteq \mathcal{F}_S(\mathcal{A})$.

In the logic ATL path quantifiers are parameterized by sets of players from $\Lambda$. The operator $\langle\langle\ \rangle\rangle$ is a path quantifier, and $\circ$ (next), $\Diamond$ (future), $\square$ (always) and $U$ (until) are temporal operators. A formula $\langle\langle\mathcal{A}\rangle\rangle\ \varphi$ expresses that the team $\mathcal{A}$ has a collective strategy to enforce $\varphi$ [9]. Boolean connectives can be defined from $\neg$ and $\vee$ in the usual way. The ATL formula $\langle\langle\mathcal{A}\rangle\rangle\ \Diamond\ \varphi$ is equivalent with $\langle\langle\mathcal{A}\rangle\rangle\ \tau\ U\ \varphi$. The constants $\tau$ and $\perp$ means *true* and *false*, respectively.

Consider a game structure $S=\langle\Lambda,Q,\Gamma,\gamma,M,d,\delta\rangle$ with $\Lambda=\{1,\ldots,k\}$ the set of players.

A *strategy* for player $a \in \Lambda$ is a function $f_a: Q^+ \to D_a$ that maps every nonempty finite state sequence $\lambda = q_0,\ldots q_n$ ($n \geq 0$), to a move of player $a$ denoted by $f_a(\lambda) \in D_a \subseteq M$. Thus, the strategy $f_a$ determines for every finite prefix $\lambda$ of a computation a move $f_a(\lambda)$ for player $a$ in the last state of $\lambda$.

Given a set $\mathcal{A} \subseteq \{1,\ldots,k\}$ of players, the set of all strategies of players from $\mathcal{A}$ is denoted by $F_{\mathcal{A}} = \{f_a \mid a \in \mathcal{A}\}$. The *outcome* of $F_{\mathcal{A}}$ is defined as $out_{F_{\mathcal{A}}}: Q \to \mathcal{P}(Q^+_\infty)$, where $out_{F_{\mathcal{A}}}(q)$ represents $q$-*computations* that the players from $\mathcal{A}$ are enforcing when they follow the strategies from $F_{\mathcal{A}}$ and $Q^+_\infty$ denotes the set of nonempty and infinite sequences over $Q$. In the following, for $out_{F_{\mathcal{A}}}(q)$ we will use the notation $out(q, F_{\mathcal{A}})$. A computation $\lambda = q_0, q_1, q_2, \ldots$ is in $out(q, F_{\mathcal{A}})$ if $q_0 = q$ and for all positions $i \geq 0$, there is a move vector $\langle j_1, \ldots, j_k \rangle \in D(q_i)$ such that:
- $j_a = f_a(\lambda[0,i])$ for all players $a \in \mathcal{A}$, and
- $\delta(q_i, j_1, \ldots, j_k) = q_{i+1}$ cf. [1].

For a game structure $S$, we write $q \vDash \varphi$ to indicate that the formula $\varphi$ is satisfied in the state $q$ of the structure $S$. An equivalent notation is $(S,q) \vDash \varphi$. The formula $\tau$ is satisfied in all states of the structure $S$ and the formula $\perp$ is not satisfied in any state of the structure $S$.

For each state $q$ of $S$, the satisfaction relation $\vDash$ is defined inductively as follows:
- for $p \in \Gamma$, $q \vDash p \Leftrightarrow p \in \gamma(q)$
- $q \vDash \neg\varphi \Leftrightarrow q \nvDash \varphi$
- $q \vDash \varphi_1 \vee \varphi_2 \Leftrightarrow q \vDash \varphi_1$ or $q \vDash \varphi_2$
- $q \vDash \langle\langle\mathcal{A}\rangle\rangle \circ \varphi \Leftrightarrow$ there exists a set $F_{\mathcal{A}}$ of strategies, such that for all computations $\lambda \in out(q, F_{\mathcal{A}})$, we have $\lambda[1] \vDash \varphi$ (the formula $\varphi$ is satisfied in the successor of $q$ within computation $\lambda$).
- $q \vDash \langle\langle\mathcal{A}\rangle\rangle \square \varphi \Leftrightarrow$ there exists a set $F_{\mathcal{A}}$ of strategies, such that for all computations $\lambda \in out(q, F_{\mathcal{A}})$, and all positions $i \geq 0$, we have $\lambda[i] \vDash \varphi$ (the formula $\varphi$ is satisfied in all states of computation $\lambda$).
- $q \vDash \langle\langle\mathcal{A}\rangle\rangle \Diamond \varphi \Leftrightarrow$ there exists a set $F_{\mathcal{A}}$ of strategies, such that for all computations $\lambda \in out(q, F_{\mathcal{A}})$ there exists a position $i \geq 0$ such that $\lambda[i] \vDash \varphi$ (the formula $\varphi$ is satisfied in at least one state of computation $\lambda$).
- $q \vDash \langle\langle\mathcal{A}\rangle\rangle \varphi_1 U \varphi_2 \Leftrightarrow$ there exists a set $F_{\mathcal{A}}$ of strategies, such that for all computations $\lambda \in out(q, F_{\mathcal{A}})$, there exists a position $i \geq 0$ such that $\lambda[i] \vDash \varphi_2$ and for all positions $0 \leq j < i$, we have $\lambda[j] \vDash \varphi_1$.

The path quantifiers $A$, $E$ of CTL can be expressed in ATL with $\langle\langle\emptyset\rangle\rangle$ and $\langle\langle\Lambda\rangle\rangle$ respectively. As a consequence, the CTL duality axioms can be rewritten in ATL, and become validities in the basic semantics: $\langle\langle\emptyset\rangle\rangle \square\ \varphi \equiv \neg\langle\langle\Lambda\rangle\rangle\Diamond\ \neg\varphi$, $\langle\langle\emptyset\rangle\rangle\Diamond\ \varphi \equiv \neg\langle\langle\Lambda\rangle\rangle\square\neg\varphi$, where the $\Lambda \in \{1,\ldots,k\}$ describe the set of players.

*Example*. In this example, adapted from [1], we will use our own definition of concurrent game structure from section 2.1, which allows the selection of some suggestive symbols for moves instead of natural numbers as in the original definition given in [1]. We consider a two-process system, $P_x$ and $P_y$. The $P_x$ process assigns boolean values to the variable $x$. When $x = \textit{false}$, then $P_x$ has two possibilities: to leave the value of $x$ unchanged or to change it in *true*. When $x = \textit{true}$, then $P_x$ has only one possibility, leaving the value of $x$ unchanged. The $P_y$ process assigns values to the boolean variable $y$ in the same way that the $P_x$ process assigns values to the variable $x$. The model of the synchronous composition of the two processes is represented by the following concurrent game structure $S=\langle \Lambda, Q, \Gamma, \gamma, M, d, \delta \rangle$:

- $\Lambda = \{1, 2\}$. Player 1 represents the process $P_x$, and player 2 represents the process $P_y$.
- $Q = \{q_0, q_1, q_2, q_3\}$. The state $q_0$ corresponds to $x = y = false$, the state $q_1$ corresponds to $x = true$, $y = false$, the state $q_2$ corresponds to $x = false$, $y = true$ and the state $q_3$ corresponds to $x = y = true$.
- $\Gamma = \{x, y\}$.
- $\gamma(q_0) = \emptyset$, $\gamma(q_1) = \{x\}$, $\gamma(q_2) = \{y\}$, $\gamma(q_3) = \{x, y\}$.
- $M = \{L, C\}$. In each state, the move $L$ of the player 1 *Leaves* the value of $x$ unchanged and the move $C$ *Changes* the value of $x$. Similarly, the move $L$ of the player 2 leaves the value of $y$ unchanged and the move $C$ changes the value of $y$ for every state of the model.
- $d_1(q_0) = \{L,C\}$, $d_1(q_1) = \{L\}$, $d_1(q_2) = \{L,C\}$, $d_1(q_3) = \{L\}$;
  $d_2(q_0) = \{L,C\}$, $d_2(q_1) = \{L,C\}$, $d_2(q_2) = \{L\}$, $d_2(q_3) = \{L\}$.
- State $q_0$ has four successors: $\delta(q_0,C,L) = q_1$, $\delta(q_0,C,C) = q_3$, $\delta(q_0,L,C) = q_2$ and respectively $\delta(q_0,L,L) = q_0$;
  State $q_1$ has two successors: $\delta(q_1,L,C) = q_3$, $\delta(q_1,L,L) = q_1$;
  State $q_2$ has two successors: $\delta(q_2,C,L) = q_3$, $\delta(q_2,L,L) = q_2$;
  State $q_3$ has one successor: $\delta(q_3,L,L) = q_3$.

The graphical representation of the concurrent game structure described above is shown in the Figure 1:

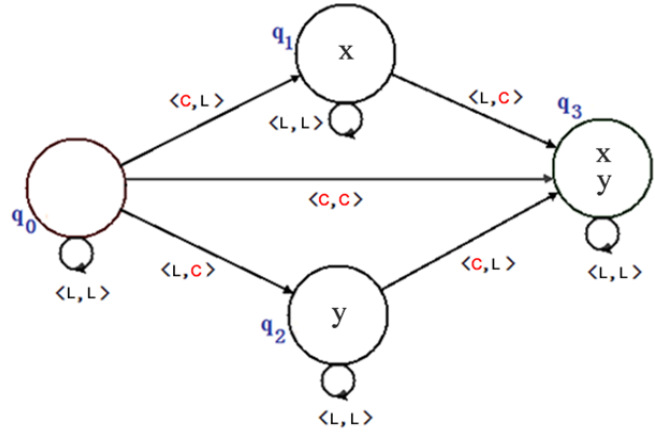

Fig. 1: A concurrent game structure example

If we consider $\mathcal{A} = \{1\}$ then we have $q_2 \vDash \langle\langle \mathcal{A} \rangle\rangle \circ (x \text{ and } y)$ and $q_3 \vDash \langle\langle \mathcal{A} \rangle\rangle \circ (x \text{ and } y)$ but $q_1 \nvDash \langle\langle \mathcal{A} \rangle\rangle \circ (x \text{ and } y)$. $q_2$ and $q_3$ are the states from which the player 1 can ensure that the immediate successor state satisfies formula ($x$ and $y$). The strategy of the player 1 is to choose the move $C$ in state $q_2$ and the player 2 does not have a strategy to avoid ($x$ and $y$) in the next state (it has only one available move, namely $L$). From state $q_1$ the player 1 does not have a strategy to enforce ($x$ and $y$) in the next state because the player 2 can choose the move $L$ and thus the successor state is also $q_1$ where the formula ($x$ and $y$) is not satisfied.

## 2.3 Formal specification of the ATL model checker tool

An ATL model checker tool consists of an algebraic compiler $\mathcal{C}:L_{atl} \to L_S$ where the source language is the language of ATL formulas and the target language is the language that describes the sets of nodes (states) of ATL models (represented by concurrent game structures) in which these formulas are satisfied. For an ATL formula $f \in L_{atl}$, $\mathcal{C}(f)=Q'$ where $Q'=\{ q \in Q \mid (S,q) \vDash f \}$.

In other words, the algebraic compiler receives as input syntactic constructions of the source language $w \in L_{atl}$ which then it maps to syntactic constructions of the target language $\mathcal{C}(w) \in L_S$.

$\mathcal{C}$ is generated from the specifications that define the model checker as a generalized homomorphism between the algebra of ATL formulas and the algebra of the set of states of the model [16]. When the homomorphism is evaluated using as input an object of the source algebra (an ATL formula), the *derived operations* are evaluated to generate the target image of the respective formula into destination algebra, the obtained result being the set of states in which formula is satisfied.

In general, a derived operation is a computation associated with an operation of the source language and is specified using syntactic constructions of the target language. The need to use derivative operations comes from the fact that often, the operations and the elements provided by the target language algebra are not expressive enough to specify the correct translation which should be performed by the algebraic compiler.

For each function name *op* from the operator scheme of algebra of the source language $L_{atl}$ a specification rule is created as a pair $(op, d_{SC}(op))$, where $d_{SC}(op)$ denote the derived operation in the syntax algebra of the target language $L_S$ which allows building the target images of expressions created in the source language by the *op*.

We denote by $O_{atl}$ the finite set of names of the operators of the language $L_{atl}$, and define it as $O_{atl} = \{ \top, \bot, \neg, \wedge, \vee, \to, \Diamond, \circ, \Box, U \}$ where $\top, \bot$ are operators of arity 0.

The set of pairs $\{ (op, d_{SC}(op)) \mid op \in O_{atl} \}$ represents the *compiler specification* that can be used to generate a compiler that will associate the words from the syntax algebra of the source language $L_{atl}$ with words from the syntax algebra of the target language $L_S$.

Implementation of the algebraic compiler $\mathcal{C}:L_{atl} \to L_S$ which represents the ATL model checker and practical performs the checking of the ATL formulas, can be synthesized by the recursive function described in Figure 2.

For formula $f$, the function $\mathcal{A}_L$ determines if it belongs to the set of generators of the $L_{atl}$ language. If $f \in \{\top, \bot\} \cup \Gamma$, $\mathcal{A}_L(f)$ returns **true**, else the function returns **false**. $\mathcal{A}_S$ is a mechanism that determines the operation and subformulas which were used to create the formula $f$.

```
function C (f ∈ L_atl)  {
    if (A_L (f)) {
        if (f = T)   return S;
        else if (f = ⊥) return ∅;
        else return { q∈Q | f ∈ γ(q) };
    }
    else if (A_S(f) = (op, (f_1,…, f_n)))
        return d_SC(op) ( C(f_1), ... , C(f_n))
    else return error;
}
```

Fig. 2: The algebraic compiler as recursive function

The components $\mathcal{A}_L$ and $\mathcal{A}_S$ of the algebraic compiler $\mathcal{C}$ can be implemented by a lexical analyzer and by a parser, respectively.

The lexical analyser $\mathcal{A}_L$ should identify the lexical atoms represented by atomic sentences correctly constructed, according to a regular grammar that generates the specification language of atomic propositions.

The parser $\mathcal{A}_S$ determines whether the formula used as input for the ATL model checker is properly constructed and belongs to the language of ATL formulas whose syntax is described using the formalism of context-free grammars.

The parser $\mathcal{A}_S$ builds the derivation tree (parse tree) of the formula into the respective grammar and thus it can determine for any sub-formula of the given formula the operation and sub-formulas used in its construction.

For the implementation of our own ATL model checker, we exploited the technological resources provided by the ANTLR system, which enables writing the derivative operations in the native language in which the whole algebraic compiler was generated (Java, C#, etc.).

To achieve the $\mathcal{C}$ compiler, we designed the following context-free grammar that generates the $L_{atl}$ language (grammar of ATL expressions) and we used ANTLR to automatically generate the components $\mathcal{A}_L$ and $\mathcal{A}_S$ on the basis of this grammar.

```
grammar ATL;

options {backtrack=true;}

@header {
   package atl;
   import java.util.HashMap;
   import org.antlr.runtime.*;
   import java.util.HashSet;
   import java.util.Iterator;
   import org.graphstream.graph.*;
   import org.graphstream.graph.implementations.*;
   }

@lexer::header {package atl;}

atlFormula
   :
   '<<A>> ' implExpr 'U' implExpr
   |       '<<A>>~' implExpr

   |       '<<A>>@' implExpr

   |       '<<A>>#' implExpr

   |       implExpr;
```

```
implExpr
 :        orExpr
          ( '=>' orExpr )* ;

orExpr
 :        andExpr
          ( 'or' andExpr )* ;

andExpr
 :        notExpr
          ( 'and' notExpr )* ;

notExpr
 :     'not'  atomExp
 |     atomExp;

atomExp
 :     '(' atlFormula ')'
 |      AP
 |      'true'
 |      'false';
```

Fig. 3: The grammar of the ATL formulas language

We note that the precedence of ATL operators is explicitly encoded by the structure of production rules.

Grammar from the Figure 3 does not contain (yet) the code necessary to implement derivative operations associated with ATL operators. From the grammar specification is observed that each ATL operator $op \in O_{atl}$ has associated a production rule.

If for the production rule $r$ we denote by $op(r) \in O_{atl}$ the ATL operator for which was defined the production $r$, a concise specification of the algebraic compiler $\mathcal{C}$ is given by the set $\{\langle r, d_{SC}(op(r))\rangle | r \in P\}$, where $d_{SC}(op(r))$ represents the derived operation in the syntax algebra of the language of states, corresponding to the ATL operator $op(r)$ and $P$ denotes the set of grammar production rules.

In the ANTLR terminology, for $d_{SC}(op(r))$ we will use the term "*semantic action attached to the production r*".

Evaluation of ATL formulas will be accomplished through implementation of the derived operations as actions attached to the production rules. Such action can be called *semantic action*, because if by example the action is attached to the production:

$$v \to v_1 v_2 ... v_n$$

the role of respective action is to calculate the semantic value of derivation subtree having the root $v$, i.e. of ATL subformula that can be built from the derivation of the nonterminal $v$.

In order to implement the compiler $\mathcal{C}: L_{atl} \to L_S$ described in Figure 2, we will transform the ATL grammar into an attribute grammar, by augmenting its production rules with semantic actions.

We present in the following a formal description of attribute grammars and the concrete use of such a grammar in implementation of the ATL model checker using ANTLR.

## 2.4 Verification of ATL models through attribute grammars

An attribute grammar may be informally defined as a context-free grammar that has been extended to provide context sensitivity using a set of attributes and evaluation rules. Each symbol (terminal or nonterminal) of an attribute grammar has associated a set (possibly empty) of attributes and each attribute has a range of possible values.

Every grammar production has an evaluation rule (or semantic rule) with actions (e.g. assignments) to modify values of attributes of symbols.

Let $G=(N,T,P,S_0)$ a context free grammar, where $N$ is the set of nonterminal symbols, $T$ – the set of terminal symbols, $P$ – the set of production rules and $S_0$ – the start symbol of the grammar.

A production $r \in P$ is of the form: $X_0 \to X_1 X_2 ... X_{n_p}$ where $n_p \geq 1$, $X_0 \in N$ and $X_k \in N \cup T$ for $1 \leq k \leq n_p$. The derivation tree of a sequence from the language generated by the grammar has the following properties:
- Each leaf node is labelled with a terminal symbol from the set $T$;
- Each inner node $t$ corresponds to a production $r \in P$, and if the production is of the form $X_0 \to X_1 X_2 ... X_{n_p}$, with the meaning of the symbols described above, then $t$ is labelled with the symbol $X_0$ and has $n_p$ child nodes labelled with $X_1, X_2, ..., X_{n_p}$ from the left to the right.

We denote by $G_A=(N,T,P,S_0,A,as)$ an attribute grammar built on grammar $G$ by its augmenting with attributes *(A)* and semantic rules/actions *(as)*. The parse tree described above is augmented with attributes of the symbols, so an attribute may have several occurrences in such a tree.

An occurrence of an attribute corresponding to a nonterminal symbol is considered evaluated when its value has been set by the semantic action(s) attached to production rule used for the nonterminal symbol in the parse tree. Occurrences of attributes corresponding to terminal symbols are evaluated by semantic actions attached to production rules used for parent nodes.

Based on the way occurrences of the attributes get their values, attributes of any nonterminal symbol $X \in N$ can be broadly divided into two categories:
- *Synthesized attributes* if their values are computed using attributes of child nodes.
- *Inherited attributes* if *their* values are computed using the values of attributes attached to the parent or siblings nodes.

We note that values of the occurrences of the same attribute may be different.

For a parse tree, the value of the designated attribute at the root can be used to associate a value, e.g. the meaning of the frontier of the tree. In our case, the frontier of a tree is a well-constructed ATL formula and the the value of the attribute at the root node represents the set of states where the ATL formula is satisfied.

In the following we will consider only *S*-attributed grammars. An attribute grammar is *S*-attributed if nonterminals have only synthesized attributes.

The set of all attributes of nonterminal $X$ is denoted by *A(X)*.

Considering the set of production rules of the form $P = \{r_1, ..., r_n\}$, with $n \geq 1$, we denote by *as(j)* = $\{action_j^{i_1}(), ..., action_j^{i_j}()\}$ the set of semantic actions attached to the production $r_j$, for each $j \in \{1, ..., n\}$. Each semantic action of a rule is implemented as a subroutine which describes the process of evaluation of any occurrence of each synthesized attribute of the nonterminal rewritten by the rule in a parse tree. For this reason, we adopt for semantic actions the syntax established in programming languages for routines and functions.

With these notations, the set of semantic actions of the grammar $G_A$ is defined as:

$$as = \bigcup_{1 \leq j \leq n} as(j)$$

The values of all occurrences of attributes of grammar symbols that appear in the derivation tree of a sequence from the language generated by the grammar are determined by the semantic actions associated with productions of the grammar that are involved in the process of derivation of the respective sequence.

These values are effectively calculated in the process of analysis, through invoking the semantic actions by the parser, generally at the moment of recognition of the next production rule used in the derivation process.

Synthesized attributes hold values used by the parent node and flow upwards so the evaluation of occurrences of synthesized attributes is done in the bottom-up manner.

We consider that ATL model is given in the form of a *concurrent game structure*: $S=\langle \Lambda, Q, \Gamma, \gamma, M, d, \delta \rangle$. The *S*-attributed grammar used in the implementation of the ATL model checker will be denoted by $G_{A,S}^{ATL} = (N, T, P, S_0, A, as)$ and has the following features:

- All attributes of nonterminal symbols $\{A(X) \mid X \in N\}$ are *synthesized attributes*;
- $|A(X)| = 1, \forall X \in N$ – any nonterminal symbol $X$ has a single attribute, denoted by $a(X)$;
- $|as(j)| = 1$, for each $j \in \{1, \ldots, n\}$ – each production rule has attached a single semantic action, $as(j) = \{action_j()\}$;
- If instead of the symbols $\top, \bot$ we are using *true* and *false*, respectively, then $T = \Gamma \cup \{true, false\}$ (the atomic propositions are terminal symbols of the *S*-attributive grammar);
- Each atomic proposition $p \in \Gamma$ has associated a single attribute denoted by $a(p)$, such as $|A(p)| = 1 \, \forall \, p \in \Gamma \subseteq T$. The set of attributes of the ATL grammar is extended to:

$$A = \{a(X)\} \mid X \in N\} \cup \{a(p) \mid p \in \Gamma\}$$

Also, each atomic proposition $p \in \Gamma \subseteq T$ has associated the semantic action $action_p()$ with the purpose of calculating the value of attribute $a(p)$. The set of semantic actions of the ATL attribute grammar becomes:

$$as = (\bigcup_{1 \leq j \leq n} \{action_j()\}) \cup (\bigcup_{p \in T} \{action_p()\})$$

- For any symbol $x \in N \cup \Gamma \cup \{true, false\}$, we denote by $v(x)$ the value of attribute $a(x)$ and we have $v(x) \subseteq Q$ (the value of the respective attribute is a subset of the set of states of the model *S*).
- If the *j*th rule, for some $1 \leq j \leq n$, was used for an occurrence of a nonterminal *X* in the parse tree, then the value of that occurrence of the attribute $a(X)$ is:

$$v(X) = action_j()$$

- The evaluation of terminal symbols is done as follows:
$v(p) = action_p() = \{q \in Q \mid p \in \gamma(q)\} \, \forall \, p \in \Gamma \subseteq T$, $v(true) = Q$ and $v(false) = \emptyset$.

Because often the model *S* is implicit, we write $G_A^{ATL}$ instead of $G_{A,S}^{ATL}$.

An ATL formula *f* is syntactically correct if and only if there is a derivation:

$$atlFormula \stackrel{*}{\Rightarrow} f$$

where we suppose that *atlFormula* = $S_0$ (the start symbol of the $G_A^{ATL}$ grammar ). The derivation tree of the formula *f* has its border composed by terminal symbols that appear in the formula *f*.

We consider that derivation $atlFormula \overset{*}{\Rightarrow} f$ has a length $k$. Then there is the sequence of direct derivations $f_0 \overset{i_1}{\Rightarrow} f_1 \overset{i_2}{\Rightarrow} f_2 \ldots \overset{i_k}{\Rightarrow} f_k$, where $f_0 = atlFormula$, $f_k = f$ and $i_l$ represents the number of production rule involved in the direct derivation $l$, where $l \in \{1,...,k\}$.

If we denote by $action_j()$ the name of the semantic action attached to the production $j \in \{1,...,n\}$, the parser will call, simultaneously with building the tree analysis of the formula $f$, the semantic actions attached to production rules from the derivation:

$$atlFormula \overset{*}{\Rightarrow} f$$

in the following order:

$$action_{i_1}(), \ action_{i_2}(), \ \ldots, \ action_{i_k}().$$

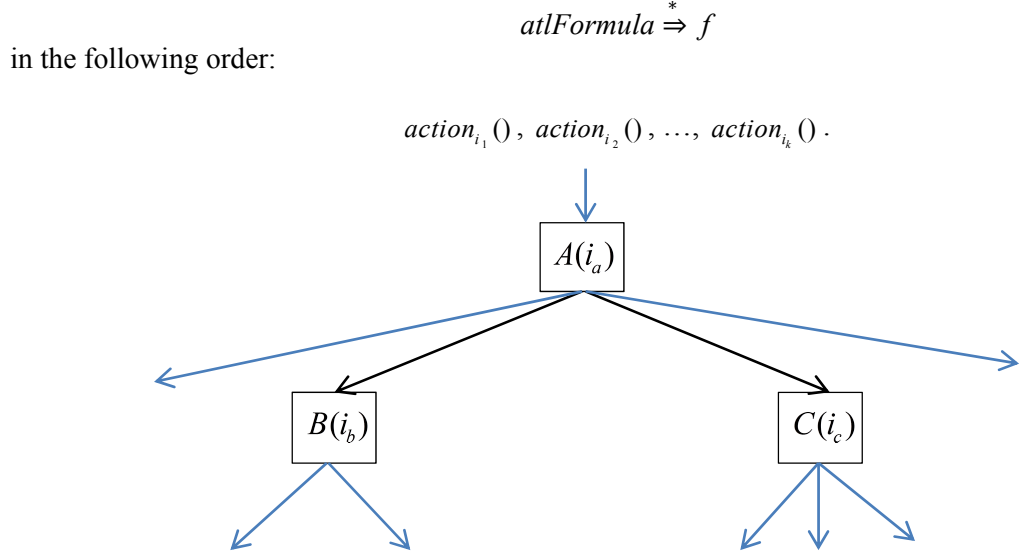

Fig. 4: A syntax subtree for the formula $f$

Assuming that in Figure 4 is represented a subtree of the derivation tree built for the formula $f$, where $A$, $B$, $C$ are nonterminals of the grammar and $i_a$, $i_b$, $i_c$ are the numbers of the production rules used in rewriting of the respective nonterminals, the function $action_{i_a}()$ will contain calls of actions $action_{i_b}()$ and respectively $action_{i_c}()$. An action will return without a call to another action only when it is attached to a terminal symbol. In this case, the respective action carries out the evaluation of that symbol. In the case of the ATL model $S=\langle \Lambda, Q, \Gamma, \gamma, M, d, \delta \rangle$, for the symbol *true* the whole set $Q$ of states will be returned. For *false* the empty set $\varnothing$ will be returned, and for some symbol $p$ from the set $\Gamma$ will be returned the set of states $\{q \in Q \mid p \in \gamma(q)\}$.

For a given formula $f \in L_{atl}$, in the function $\mathcal{C}(f)$ the parser $\mathcal{A}_s$ identifies the first production of grammar used in derivation:

$$S_0 \overset{*}{\Rightarrow} f$$

Assuming that production is $r$: $S_0 \rightarrow t_0 X_1 t_1 X_2 ... X_m t_m$, $\mathcal{C}(f)$ will call the derived operation associated with production $r$, $d_{SC}(op(r))$, and will store the result in the meta-variable $set$, as follows:

$$\$set = d_{SC}(op(r))(\$a_1.set, \ldots \$a_m.set)$$

where $\$a_i.set$ are semantic evaluations of nonterminals $X_i$, $1 \leq i \leq m$, by whose rewriting are obtained the subformulas $f_1, \ldots, f_m$ of $f$. These evaluations are performed recursively, and we have:

$$\$a_i.set = \mathcal{C}(f_i), \ 1 \leq i \leq m$$

If we denote by $G=(N,T,P,S_0)$ the context-free grammar that generates the ATL formulas language, with production set in the form of $P = \{r_1, \ldots, r_n\}$, $n \geq 1$, a concise specification of the compiler $\mathcal{C}$ is given by the set of pairs $\{\langle r_i, d_{SC}(op(r_i)) \rangle | 1 \leq i \leq n \}$, where $d_{SC}(op(r_i))$ is the *derived operation* corresponding to the production $r_i$ and $op(r_i)$ is the ATL operator for which was defined the production $r_i$, $1 \leq i \leq n$.

Automatic generation of the ATL model checker from the above specification is accomplished in ANTLR by building an attribute grammar $G_A^{ATL} = (N,T,P,S_0,A,as)$ in the meta-description language of ANTLR grammars, with the following properties:

- The grammar productions are those specified in the Section 2.3.
- Attributes associated to the generators of language $L_{atl}$ have the following values:

$$v(p) = action_p() = \{q \in Q | p \in \gamma(q)\}, \forall p \in \Gamma \subseteq T$$
$$v(true) = action_{true}() = Q$$
$$v(false) = action_{false}() = \varnothing$$

- For production $r_i$: $X_0 \to t_0 X_1 t_1 ... X_m t_m$, the attribute value of nonterminal $X_0$ is calculated as:

$$v(X_0) = action_i() = d_{SC}(op(r_i))(v(X_1),...,v(X_m)).$$

The attribute flow algorithm propagates attribute values through the parse tree by traversing the tree according to the *set* and *used* dependencies between attributes (an attribute must be *set* before it is *read*). The values $v(X_1),...,v(X_m)$ must be known before being used in the evaluation of the attribute of nonterminal $X_0$. The evaluation of the attributes of nonterminals $X_1,...,X_m$ is done in the body of the function $action_i()$ and is explicitly suggested by the construction $d_{SC}(op(r_i))(v(X_1),...,v(X_m))$. We can consider that the call of derived operation $d_{SC}(op(r_i))(v(X_1),...,v(X_m))$ is performed inside of the function $action_i()$.

Using the notational descriptions of the meta-description language used in specification of the ANTLR attribute grammars, equality becomes:

$$\$set = d_{SC}(op(r_i))(\$a_1.set ,..., \$a_m.set).$$

The components $\mathcal{A}_L$, $\mathcal{A}_S$ of the algebraic compiler are automatically generated by ANTLR, using as input the unique file containing the definition of the grammar $G_A^{ATL}$.

In the ANTLR grammar, the meta-variables of form $set are used to store the attribute values of grammars symbols ($N \cup T$). At the time of construction of the derivation tree, for meta-variables that appear in the definition of a semantic action attached to a production used in the derivation process, ANTLR generates code to invoke the semantic actions of productions used in rewriting of nonterminals appearing in the right member of the corresponding production.

The role of semantic actions associated with the production rules of the ANTLR grammar is to calculate and return the attribute values of nonterminal symbols from the left member of respective productions (nonterminals rewritten by these productions).

For example, for the rule *atomExp*:

```
atomExp returns [HashSet set]
:
...
;
```

the generated code has the following form (simplified):

```
public HashSet atomExp () throws RecognitionException
{
      HashSet set = null;   // Return value, referenced in
      ...                   // the definition file of grammar
      return set;           // by $set
}
```

# 3 Calculation

Verification of the given formula $f$ involves the building of the derivation $atlFormula \stackrel{*}{\Rightarrow} f$ (and hence the corresponding derivation tree) and it is fully completed when the attribute value of the start symbol of the grammar is computed. This value, $v(atlFormula) \subseteq Q$, represents the set of all states from $Q$ which satisfy the formula $f$ in the given model $S$.

Building a model checker based on an attributive grammar requires a detailed description of its semantic actions.

The implementation of the ATL temporal operators is based on function $Pre(\mathcal{A}, \Theta)$ – the set of states from which players $\mathcal{A}$ can enforce the system into some state in $\Theta$ in one move, where $\mathcal{A} \subseteq \Lambda$ and $\Theta \subseteq Q$. In other words $Pre(\mathcal{A}, \Theta) = \{q \in Q \mid \forall a \in \mathcal{A} \ \exists j_a \in d_a(q)$ such that for every $b \in \Lambda \setminus \mathcal{A}$ and $j_b \in d_b(q)$, $\delta(q, j_1, \ldots, j_k) \in \Theta\}$.

In case of large ATL models, with many states and players, it is very important for the model checker tool to have an efficient implementation for $Pre(\mathcal{A}, \Theta)$ function. In the following, we describe an original implementation of the $Pre()$ function using relational databases [17].

For a concurrent game structure $S$ presented in section 2.1, can be defined a directed multi-graph $G_S = (X, U)$, where $X = Q$, and $(b, e) \in U \Leftrightarrow \exists \langle j_1, \ldots, j_k \rangle \in D(b)$ such as $\delta(b, j_1, \ldots, j_k) = e$. The labelling function for the graph $G_S$ is defined as follows: $L: U \to \mathcal{P}(\overline{M})$, $\forall u = (b, e) \in U$, $L(u) = \{\langle j_1, \ldots, j_k \rangle,$ where $\delta(b, j_1, \ldots, j_k) = e\}$.

We denote by $MOVES_\Lambda = \bigcup_{q \in Q} D(q) = \{\langle j_1, \ldots, j_k \rangle \mid j_a \in D_a, \forall a \in \Lambda\}$ the set of all move vectors within the concurrent game structure $S$. Because $MOVES_\Lambda$ is a set of tuples, it can be seen as a relation having as attributes the set $\Lambda$, each attribute $a \in \Lambda$ having as a set of possible values $D_a$ (cf. (1) from section 2.1).

For a given set $\mathcal{A} = \{i_1, \ldots, i_n\}$ of $n$ players, the projection $\pi_\mathcal{A}(MOVES_\Lambda)$ eliminates all attributes of the input relation excepting those mentioned in the set $\mathcal{A}$ and produces for each input tuple $\langle j_1, \ldots, j_k \rangle$ an output tuple $\langle j_{i_1}, \ldots, j_{i_n} \rangle$.

With the notation $\pi_\mathcal{A}(MOVES_\Lambda) = MOVES_\mathcal{A}$, for the set $\mathcal{A} = \{i_1, \ldots, i_n\}$ of $n$ players we define $L_\mathcal{A}: U \to MOVES_\mathcal{A}$, $L_\mathcal{A}(u) = \{\langle j_{i_1}, \ldots, j_{i_n} \rangle \mid j_{i_a} \in \{j_1, \ldots, j_k\} \ \forall a \in \{1, \ldots, n\}$, where $u = (b, e)$ and $\delta(b, j_1, \ldots, j_k) = e\}$.

We create a table *model* in a database in order to represent each $u = (b, e) \in U$ having three columns: a column $B$ which contains the value $b$, a column $E$ which contains the value $e$ and a column LABEL which contains the values $L_\mathcal{A}(u)$ in distinct rows.

Using this notations, our implementation of the $Pre(\mathcal{A}, \Theta)$ function using SQL statements is presented in the Figure 5:

```
select distinct B from
(
    select distinct x.B, y.LABEL from
    (
        select distinct B, LABEL from model
        where E in Θ
    ) x
    left join
    (
        select distinct B, LABEL from model
        where E not in Θ
    ) y
    on x.B = y.B and x.LABEL = y.LABEL
    where y.LABEL is null
) z
```

Fig. 5: The implementation of the $Pre(\mathcal{A}, \Theta)$ function using SQL statements

The function $Pre(\mathcal{A}, \Theta)$ is implemented in Java code, specifications being included in the single definition file of the ATL grammar. The function $Pre(\mathcal{A}, \Theta)$ is dependent on the verified ATL model, so the model must be accessible to the algebraic compiler when verifying an ATL formula, in the form of internal data structures required by the call.

The advantage of this solution is that entire algebraic compiler code is generated in a single step without the need for previous pre-processing.

For the □ ATL operator we use in ANTLR the # symbol. Also, we denote the ◊ ATL operator with ~ symbol and the ○ operator is replaced by @ symbol.

The formal definitions of the derivative operations $d_{SC}(op)$ which correspond to the temporal ATL operators are presented in Table 1:

| atlFormula: $set → '<<A>>#'<br>implExpr : $set1<br>{ Set Z, Z1;<br>   Z:=Q; Z1:= $set1;<br>   while (¬(Z ⊆ Z1)) {<br>      Z:=Z1;<br>      Z1:= Pre(𝒜,Z) ∩ $set1;<br>   }<br>   $set :=Z;<br>} | atlFormula: $set → '<<A>>'<br>implExpr:$set1 'U' implExpr:$set2<br>{ Set Z,Z1;<br>   Z:= ∅; Z1:=$set2;<br>   while(¬(Z1 ⊆ Z)) {<br>      Z:= Z ∪ Z1;<br>      Z1:=Pre(A,Z) ∩ $set1;<br>   }<br>   $set:=Z;<br>} |
|---|---|
| atlFormula:$set → '<<A>>~'<br>implExpr:$set1<br>{ Set Z,Z1;<br>   Z:= ∅; Z1:= $set1;<br>   while(¬(Z1 ⊆ Z))<br>   {<br>      Z := Z ∪ Z1;<br>      Z1:= Pre(A,Z) ∩ Q;<br>   }<br>   $set:=Z;<br>} | atlFormula:$set → '<<A>>@'<br>implExpr:$set1<br>{<br>   $set := Pre(A,$set1);<br>} |

Table 1: The formal definition of the derivative operations associated with temporal operators

The semantic action of production corresponding to the □ operator implements the derivative operation $d_{SC}(\square)$. The argument of the derived operation, $set1$, represents the calculated image (satisfaction set) of the ATL sub-formula to which is applied the □ operator.

The value returned by the semantic action is stored in the variable $set$ to be propagated in the analysis / evaluation process of the ATL formula for which the process of verification was launched:

$$\$set = d_{SC}(\square) \ (\$set_1).$$

For the ATL operator □, the corresponding *action* included in our ANTLR grammar of ATL language is detailed in Figure 6.

```
atlFormula returns [HashSet set]
@init { }
: '<<A>>#' e=implExpr
    {
        HashSet r = new HashSet(all_setS);
        HashSet p = $e.set;
        while (!p.containsAll(r))
        {
            r = new HashSet(p);
            p = Pre(r);
            p.retainAll($e.set);
        }
        $set = r;
    }
```

Fig. 6: Implementation of the □ (*always*) operator in ANTLR

In our implementation the *all_setS* is *Q*, and means all states of the model and the local variables *p*, *r* are sets used in the internal implementation of the algebraic compiler.

## 4 Experimental

In this section we evaluate the effectiveness of our approach in designing and implementing an ATL model checker and we report some experimental results.

For the beginning we describe the usage of our model-checker to design a game strategy when playing Tic-Tac-Toe (called TTT for short in the rest of this paper). Although the game implemented is relatively simple, due to the large size of the structure representing the ATL model at the first moves, it represents a good opportunity to study the impact of technologies used to implement the model checker in its performance.

It is shown in [13] that the model checking of computation tree logic (CTL) formulae can be used for generating plans in deterministic as well as non-deterministic domains. Because ATL is an extension of CTL that includes notions of agents, their abilities and strategies (conditional plans) explicitly in its models, ATL is better suited for planning, especially in multi-agent systems [6].

ATL models generalize turn-based transition trees from game theory and thus it is not difficult to encode a game in the formalism of concurrent game structures, by imposing that only one player makes a move at any given time step.

The game TTT is played by two opponents with a turn-based modality on a 3×3 board. The two players take turns to put pieces on the board. A single piece is put for each turn and a piece once put does not move. A player wins the game by first lining three of his or her pieces in a straight line, no matter horizontal, vertical or diagonal.

The implemented algorithm looks for infallible conditional plans to achieve a winning strategy that can be defined via ATL formulae.

We consider a computer program playing TTT game with a user (human) and the ATL model checking algorithm is used to return a strategy to achieve a winning strategy for the computer. The TTT is a turn-based synchronous game. In such a system, at every transition there is just one player that is permitted to make a choice (and hence determine the future).

Formally, a game structure $S=\langle\Lambda,Q,\Gamma,\gamma,M,d,\delta\rangle$ is turn-based synchronous if for every state *q* from its finite set of states *Q*, there exist a player *a* from the set of all players $\Lambda$ such that $|d_b(q)| = 1$ for all players $b\in\Lambda\setminus\{a\}$. State *q* is the *turn* of player *a*.

### 4.1 Modelling The Game

In the following we will show how to use the ATL formalizations in finding winning strategies in case of TTT game.

We transform the original problem into an ATL model checking problem. More specifically, we want to determine a strategy $f_a : Q^+ \to D_a$ which leads the game into a winning state for the player $a\in\Lambda$ representing the computer.

We suppose that positions of the board are numbered as in Figure 7:

| 0 | 1 | 2 |
| 3 | 4 | 5 |
| 6 | 7 | 8 |

Fig. 7: Labelling the grids on the board

Formally, the turn-based synchronous game structure of TTT is defined as follows: $S=\langle\Lambda,Q,\Gamma,\gamma,M,d,\delta\rangle$ where the set of players is $\Lambda =\{1,2\}$ and we consider that computer is represented by player 1 and the user is represented by the player 2.

Values of the board locations are denoted by $x_i \in \{0,1,2\}$, where $i \in \{0,1,...,8\}$. The value 0 means an empty position, the value 1 denotes a previous move of the player 1 and the value 2 represents a move of the player 2. For the sequence of values $\overline{x_l x_m x_n}$ we define $\#\overline{x_l x_m x_n} = \min(x_l,1) + \min(x_m,1) + \min(x_n,1)$ where $l, m, n \in \{0,1,...,8\}$.

For each state $q \in Q$ of the game we have:

$$\gamma(q) = \{ \overline{x_l x_{l+1} x_{l+2}}_{l=0,3,6}, \overline{x_l x_{l+3} x_{l+6}}_{l=0,1,2}, \overline{x_0 x_4 x_8}, \overline{x_2 x_4 x_6}, T \mid x_k \in \{0,1,2\} \text{ for } k \in \{0,..,8\} \text{ and } T \in \{1,2\} \}.$$

A state labelled with value $T = 1$ means that is turn of the player 1 for making the move and if the state is labelled with $T = 2$ then the player 2 will make the next move.

The set of possible movements of players is $M=\{0,1,2,3,4,5,6,7,8,9\}$.

For the player 1, the set of alternative movements in the state $q \in Q$, in case when movements are still possible, is defined as:

$$d_1(q) = \begin{cases} \{1,...,k\} \text{ if } k = 9 - \sum_{l=0,3,6} \#\overline{x_l x_{l+1} x_{l+2}} \geq 1 \text{ and } 1 \in \gamma(q) \\ \{0\} \text{ if } k = 9 - \sum_{l=0,3,6} \#\overline{x_l x_{l+1} x_{l+2}} \geq 1 \text{ and } 2 \in \gamma(q) \end{cases}$$

In a similar manner are defined the possible movements of the player 2.

The game stops (so no moves are possible) if the board moves locations are full i.e.:

$$\sum_{l=0,3,6} \#\overline{x_l x_{l+1} x_{l+2}} = 9$$

Another situation where the game is not continuing is when a player won. The state $q$ is a winning state for player 1 if $\overline{111} \in \gamma(q)$ and it is a winning state for player 2 if $\overline{222} \in \gamma(q)$.

Alternation to move can be formalized as follows: for a transition $\delta(q, j_1, j_2) = q'$, there are the following cases:

$$1 \in \gamma(q) \Rightarrow 2 \in \gamma(q') \text{ or } 2 \in \gamma(q) \Rightarrow 1 \in \gamma(q')$$

In order to win the game, the player 1 (the computer) must follow two rules:

1. Try to choose at next move a state from the set $\langle\langle 1 \rangle\rangle \Diamond (\overline{111})$, which favours the wining of the game in the future.
2. Do not choose at next move a state from the set *avoid* defined as follows:

$$avoid = \begin{cases} \langle\langle 2 \rangle\rangle \circ (\overline{222}) \text{ if the computer made the first move} \\ \langle\langle 2 \rangle\rangle \Diamond (\overline{222}) \text{ if the user made the first move} \end{cases}$$

to prevent the player 2 to win on the next move.

### 4.2 Easy to use ATL model checker tools

We choose to publish our implementation of the ATL model checker as a Web service in order to make the core of our tool accessible to various clients.
Our implementation is based on GlassFish/Tomcat as a Web container, and relies on different databases: MySQL, SQLServer and H2. ATL model checking tools implement both Web services architectures: SOAP-based (JAX-WS Web Services) and respectively RESTful (using Jersey, the open source JAX-RS Reference Implementation for building RESTful Web services).
For testing purposes, the ATL model checker described in this paper is available online via Web services hosted by a public VPS (Linux) and is accessible through the client component of the ATL suite – ATLDesigner, a C# application available at http://use-it.ro/. ATLDesigner allows interactive graphical development of the ATL models. The server components are provided as Web applications which can be easily deployed in GlassFish or Tomcat applications servers. For a local installation, the Web Services endpoints are:

- http://localhost:8080/ATL-Checker/ATLCheckerService?WSDL – SOAP version;
- http://localhost:8080/ATLChecker/service/post – RESTful version.

Details about coding a model for its forwarding to a Web service (XML or JSON format) are provided by ATLDesigner (select the button *Web Service* from the main toolbar and then click on button *Show model* from the window that opens).

For programmatic construction of huge ATL models are available ATL Libraries for Java and C# languages.

### 4.3 Experimental Results

The major impact on performance of the ATL model checker is represented by the implementation of the function *Pre()*, which was presented in detail in section 3 and is based exclusively on the database server used.

In order to analyse their impact in the performance of the ATL model checker, were used two different database servers to implement the Web service, namely MySql 5.5 and respectively H2.

In [15] the Tic-Tac-Toe was implemented in the Reactive Modules Language (RML). RML is the model description language of the ATL model checker MOCHA, which was developed by Alur et al. [2]. Experimental results showed that the time necessary to find a winning strategy for a player, on a configuration with a Dural-Core 1.8Ghz CPU, was 1 minute and 6 seconds [15]. Running on Intel I7 mobile version configuration, our ATL model checker tool is able to find a winning strategy in about 9 seconds using MySql as a database server and 7 seconds when H2 was used.

In Table 2 are presented experimental results from using our tool to get a winning strategy in the TTT game.

The best results are obtained when using the H2 (in-memory) database and the RESTful Web service. In the actual stage of the development, experimental results are encouraging, showing that our tool is able to handle large systems efficiently. By using a database-based technology in the core of the ATL model checker, our tool provides a good foundation for further improvement of its performance and scalability.

| Total time necessary to determine the winning strategy (Tic-Tac-Toe game) Intel Core I7 5500U, 2.4 GHz, 8Gb RAM | | | | |
|---|---|---|---|---|
| **Number of states** | MySQL 5.5 (seconds) | | H2 1.3 (seconds) | |
| | SOAP | RESTful | SOAP | RESTful |
| 35027 | 15.55 | 9.23 | 13.83 | 6.88 |
| 32251 | 13.53 | 8.37 | 11.66 | 6.28 |
| 31101 | 12.87 | 8.02 | 11.00 | 6.01 |
| 28817 | 11.18 | 6.96 | 9.52 | 5.20 |
| 28267 | 11.01 | 6.93 | 9.27 | 5.11 |
| 27647 | 10.35 | 6.59 | 8.95 | 4.96 |
| 26853 | 10.01 | 6.48 | 8.8 | 4.78 |
| 22851 | 7.18 | 4.68 | 6.39 | 3.66 |
| 4791 | 1.43 | 1.4 | 1.1 | 0.91 |
| 4255 | 1.22 | 1.17 | 0.86 | 0.75 |
| 3732 | 1.12 | 1.03 | 0.80 | 0.65 |
| 3683 | 0.98 | 0.93 | 0.70 | 0.62 |
| 3423 | 0.94 | 0.88 | 0.68 | 0.59 |
| 2307 | 0.68 | 0.66 | 0.48 | 0.41 |

| Total time necessary to determine the winning strategy (Tic-Tac-Toe game) Intel Core I7 5500U, 2.4 GHz, 8Gb RAM | | | | |
|---|---|---|---|---|
| **Number of states** | MySQL 5.5 (seconds) | | H2 1.3 (seconds) | |
| | SOAP | RESTful | SOAP | RESTful |
| 2236 | 0.66 | 0.65 | 0.46 | 0.39 |

Table 2: The impact of Database servers and Web Services technologies in performance of the ATL model checker

## 5 Results and Discussion

In this paper we built an ATL model checking tool, based on robust technologies (Java, .NET, SQL) and well-known standards (XML, SOAP, JSON, HTTP). The implementation of the ATL model checking algorithm is based on Java code generated by ANTLR using an original ATL grammar extended to an attribute grammar, by augmenting its production rules with semantic actions.

We enumerate some of the arguments that recommend the utilization of the ANTLR attribute grammars in implementing model checker tools:
- The verified model can be encoded and accessed by classes of objects in the chosen target language (C++, Java, C#, Objective C, Python) directly in the attribute grammar specification file.
- For the implementation of the semantic actions in ANTLR, we can exploit the full power of an advanced programming language (Java, C #, etc.).
- We can specify multiple target languages to generate the model checker tool, and the semantic actions can be implemented by efficient code, taking into account the features of the chosen language.
- The proposed methodology has a generic character since it can be applied to generate model checkers for different temporal logics (CTL, ATL, LTL, etc.).

The server components of our tool (ATL Checker) are published as Web Services, exposing their functionality through standard XML/JSON interfaces. The ATL Model Checking suite (the ATL Designer, Web Services packages and ATL API Client libraries for Java and C#) can be downloaded from http://use-it.ro.

By using a database-based technology in the core of the ATL model checker, our tool provides a good foundation for further improvement of its performance and scalability. The use of powerful database systems has a positive impact in overall performance of our ATL model checker. Also, the RESTful architecture provides the best results: it has a simpler style and is less verbose so that less volume of data is sent when communicating between the client and server.

Further investigation on improving performance will be done using in-memory databases (H2, HSQLDB, MariaDB MEMORY storage engine). Because the SQL queries used in verification a composite ATL formula might consist of many subqueries that can be run in parallel, we would start looking at using the horizontal scalability features such as Parallel Pipelined Table Functions (PTF) provided by Oracle databases.

**Acknowledgements.** Project financed from "Lucian Blaga" University of Sibiu research grants LBUS-IRG-2015-01, and developed within the Center of Scientific Research in Informatics and Information Technology.